\documentclass[pre,aps,superscriptaddress,twocolumn,floatfix]{revtex4-1}
\usepackage{dsfont}
\usepackage{mathtext}

\usepackage{mathtools}
\usepackage[usenames,dvipsnames]{xcolor}
\usepackage{amsmath}
\usepackage{amssymb,mathrsfs}
\usepackage{graphicx}
\usepackage{epstopdf}

\usepackage{mathtext}

\usepackage{bm}

\newcommand{\br}{{\bm r}}

\newcommand{\tpsi}{\tilde{\psi}} 

\newcommand{\hbi}{\hat{\bm i}}
\newcommand{\hbj}{\hat{\bm j}}

\begin{document}
	
	\title{Floquet defect solitons}

	\author{Sergey~K.~Ivanov}
	\affiliation{Moscow Institute of Physics and Technology, Institutsky lane 9, Dolgoprudny, Moscow region, 141700, Russia}
	\affiliation{Institute of Spectroscopy, Russian Academy of Sciences, Troitsk, Moscow, 108840, Russia}
	
	\author{Yaroslav~V.~Kartashov}
	\affiliation{Institute of Spectroscopy, Russian Academy of Sciences, Troitsk, Moscow, 108840, Russia}
	\affiliation{ICFO-Institut de Ciencies Fotoniques, The Barcelona Institute of Science and Technology, 08860 Castelldefels (Barcelona), Spain}
	
	\author{Vladimir~V.~Konotop}
	\affiliation{Departamento de F\'isica, Faculdade de Ci\^encias, Universidade de Lisboa, Campo Grande, Ed. C8, Lisboa 1749-016, Portugal}
	\affiliation{Centro de F\'isica Te\'orica e Computacional, Universidade de Lisboa, Campo Grande, Ed. C8, Lisboa 1749-016, Portugal}

\begin{abstract}
We consider an array of straight nonlinear waveguides constituting a two-dimensional square lattice, with a few central layers tilted with respect to the rest of the structure. It is shown that such configuration represents a line defect, in the lattice plane, which is periodically modulated along the propagation direction. In the linear limit such a system sustains line defect modes, whose number coincides with the number of tilted layers. In the presence of nonlinearity the branches of defect solitons propagating along the defect line bifurcate from each of the linear defect modes. Depending on the effective dispersion induced by the Floquet spectrum of the underline system the bifurcating solitons can be either bright or dark. Dynamics and stability of such solitons are studied numerically.
\end{abstract}

%\setboolean{displaycopyright}{true}

\maketitle
Dynamic modulations of parameters of periodic optical  systems may dramatically affect their eigenmode structure leading to a host of new physical phenomena such as dynamic localization, diffraction management, stimulated mode conversions, and formation of new defect or surface localized states~\cite{dinlattices01}. They may also introduce nontrivial topology of the Floquet bands that may be accompanied by the formation of topologically protected edge states~\cite{topophot02}. Especially interesting in this respect is a class of transversely periodic systems with modulations 
%(driving) 
that are also periodic in the evolution variable (in paraxial optics - in the propagation direction), that enables Floquet engineering of their band structures, the importance of which is also recognized in quantum mechanics \cite{floquet04,floquet02}. Such periodically driven systems offer unique opportunities for realization of artificial gauge fields for electromagnetic waves~\cite{topophot01,topophot02}, obtaining topologically nontrivial phases, and  formation of unidirectional edge states proposed in \cite{floquet05}, analyzed theoretically in \cite{floquet06}, and   observed experimentally at optical frequencies in modulated waveguide arrays \cite{floquet07,floquet08,floquet09}. Inclusion of material nonlinearities in such waveguiding systems enables the formation of topological Floquet edge solitons \cite{edgesol01,edgesol02, edgesol03,edgesol04,edgesol05,edgesol06,edgesol07,edgesol08,edgesol09,edgesol10} recently observed experimentally \cite{solobsv01,solobsv02}, and nonlinearity-induced topological Floquet insulators \cite{solobsv03} (see e.g.,~\cite{nonlrev01} for a recent review).

One of the intriguing consequences of longitudinal driving is a possibility of formation of unconventional localized defect modes at the edges of topologically trivial periodically curved waveguide arrays 
%at specific bending amplitudes 
\cite{curved01,curved02,curved03}. Periodic driving leads also to the appearance of new states of topological origin (sometimes coexisting with topologically trivial edge modes) in driven Aubry-Andr\'e-Harper \cite{drivenaah01} and Su-Schrieffer-Heeger waveguide arrays \cite{drivenssh01,drivenssh02,drivenssh03,drivenssh04, drivenssh05}. Recently, {linear} localized Floquet modes have been observed at the interfaces between groups of waveguides with different bending or tilting laws interpreted in \cite{defect01,defect02} as gauge-induced localization. Such modes appear as extended states diffracting along the tilted or differently curved layers sustaining them. {Thus, while an explanation of the linear mechanism of localization across tilted/bent layers was suggested in \cite{defect01}, the formation of hybrid self-sustained states localized also along such layers due to nonlinearity was not addressed, so far.}

\begin{figure}[t]
\centering\includegraphics[width=\linewidth]{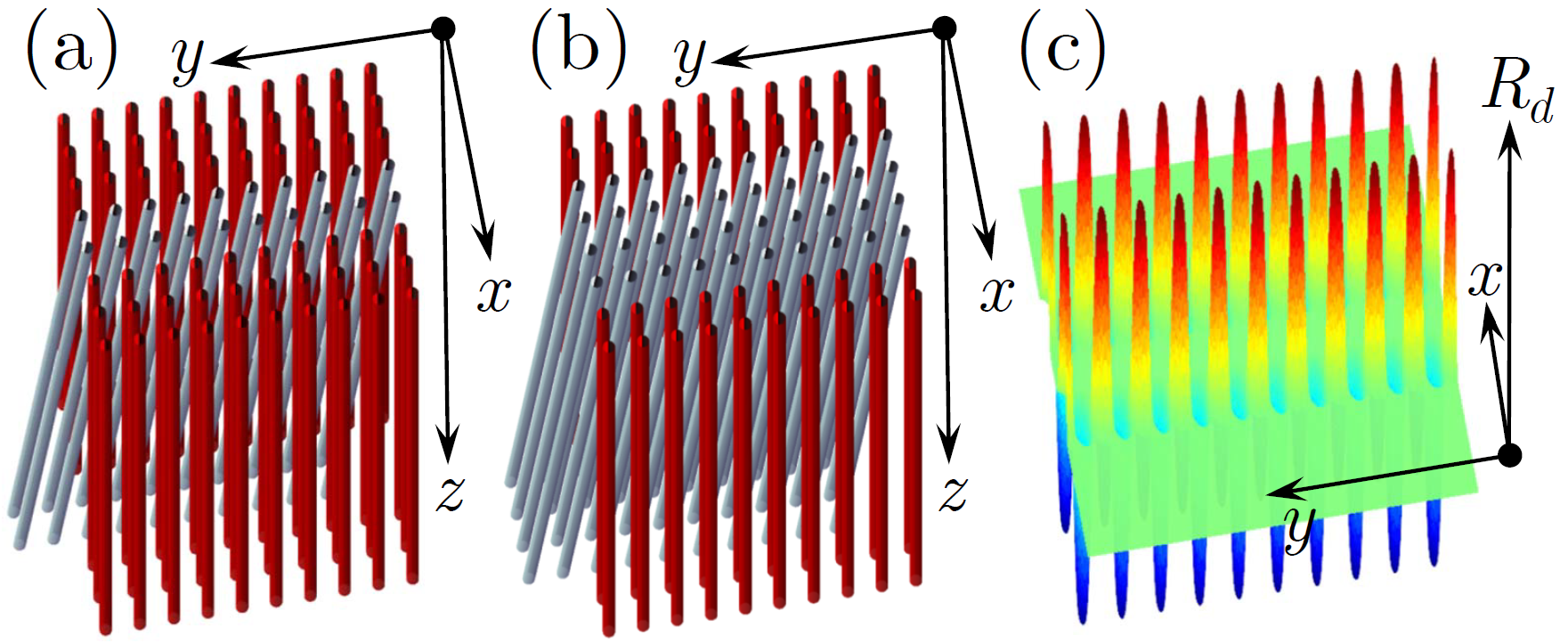}
\caption{Examples of arrays with line defects consisting of $M=2$ (a) and $M=4$ (b) tilted layers. Straight waveguides are shown red, tilted ones are shown grey. In (c) the profile of the defect potential $R_\textrm{d}$ for $M=2$ is depicted at $z=Z/2$.}
\label{figure1}
\end{figure}

The goal of the present Letter is to introduce Floquet defect solitons in longitudinally modulated nonlinear waveguide arrays. Such solitons are obtained in two-dimensional square waveguide arrays, where a few central layers are tilted (vertically) 
with respect to other straight layers in the structure, while periodicity in the vertical direction is not affected by the tilt [see schematics in Fig.~\ref{figure1}(a),(b)].  Such Floquet system supports line defect modes, {experimentally observed in \cite{defect01},} that are localized in the direction transverse to the tilt direction and are extended along the defect. We find that focusing nonlinearity of the material leads to bifurcation from a linear extended defect mode of a family of nonlinear localized modes, propagating along the tilted layer. Such {hybrid} nonlinear modes can be termed as {\em Floquet defect solitons} propagating along the defect but having nontrivial field distribution in the direction orthogonal to the defect line. One can find both bright and dark defect solitons, which are shown to be robust entities surviving over very long propagation distances.

Propagation of a paraxial light beam along the $z$ axis in a two-dimensional waveguide array with focusing cubic (Kerr) nonlinearity is described by the nonlinear Schr\"{o}dinger (NLS) equation for the dimensionless field amplitude $\psi$:
\begin{equation} 
\label{NLS_dimensionless}
    i \frac{\partial \psi}{\partial z} =  H_0\psi - |\psi|^2\psi, \quad H_0=-\frac{1}{2} \nabla^2  -\mathcal{R}(\br,z).
\end{equation}
Here $\br=x\hbi+y\hbj$ and $\nabla=(\partial_x,\partial_y)$. The transverse coordinates ${\br}$ and the propagation distance $z$ are normalized to the characteristic transverse scale $w$ and the diffraction length $\kappa w^2$, respectively,
{where $\kappa$ is the wavenumber}. Physically, the system considered here consists of straight waveguides assembled in two semi-infinite square arrays and separated by $M$ tilted mono-layers of waveguides [see schematic representations in Figs.~\ref{figure1}(a),(b) for $M=2$ and $M=4$]. The tilted waveguides have the same physical characteristics as the waveguides of the semi-infinite arrays. Mathematically, the refractive index of such system can be viewed as an ideal $z$-independent square optical potential $R_\textrm{s}(\br)$ with a lattice constant $\ell$, $R_\textrm{s}(\br) = R_\textrm{s}(\br+\ell \hbi) = R_\textrm{s}(\br+\ell \hbj)$, perturbed by a line defect. The defect is periodic along $z$ and $y$ directions, $R_\textrm{d}(\br,z)=R_\textrm{d}(\br,z+Z)=R_\textrm{d}(\br+\ell\hbj,z)$, with $Z$ being the longitudinal period, and it has zero boundary conditions at $|x|\to \infty$:  $\lim_{|x|\to\infty}R_\textrm{d}(\br,z)=0$. Respectively, the resulting optical potential of the whole system
%, $\mathcal{R}(\br,z)$, 
can be written as $\mathcal{R}(\br,z)=R_\textrm{s}(\br)+R_\textrm{d}(\br,z)$.  For the choice $R_\textrm{s}(\br)=p\sum_{m,n} \exp{\left(-[\br-\ell(m\hbi+n\hbj)]^2/a^2\right)}$, modelling the perfect reference array of waveguides of the radius $a$, the line defect created by $M$ tilted waveguides has the form $R_\textrm{d}\equiv R_M(\br-\alpha z\hbj)-R_M(\br)$, where $R_M(\br)=p\sum_n\sum_{m=1}^M\exp{\left(-[\br-\ell(m\hbi+n\hbj)]^2/a^2\right)}$ and $\alpha=\ell/Z$ is the tilt angle of $M$ defect layers with respect to the reference array. The line defect is illustrated in Fig.~\ref{figure1}(c). Note that $R_\textrm{d}(\br,nZ)=0$ ($n$ is an integer) and the array $\mathcal{R}(\br,nZ)$ becomes a uniform square lattice in the transverse plane.

%%%%%%%%%%%%%%%%%%%%%%%%%%%%%%%%%%%%%%%%%%%%%%%%%%%%%%%%
%%%%%%%%%%%%%%%%%%%%%%%%%%%%%%%%%%%%%%%%%%%%%%%%%%%%%%%%

\begin{figure}[t]
\centering\includegraphics[width=\linewidth]{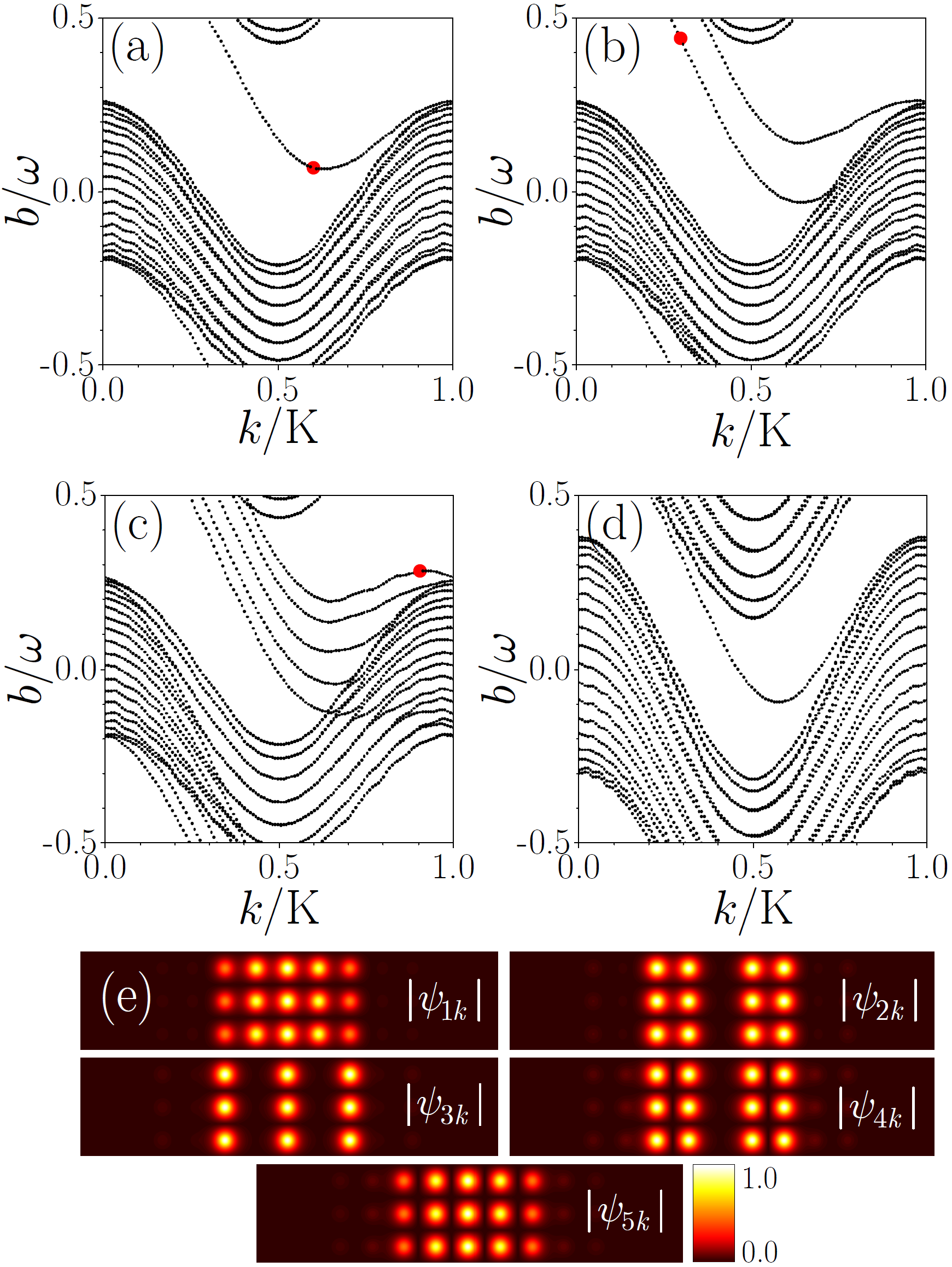}
\caption{Quasi-propagation constants versus Bloch momentum for structures containing $M=1$ (a), $M=2$ (b), and $M=5$ (c) tilted layers with longitudinal period $Z=10$, and $M=1$ layer with longitudinal period $Z=15$ (d). Red dots indicate modes on which solitons are built on. (e) Examples of linear defect modes (at $z=0$) supported by $M=5$ tilted layers at $k_0=0.55K$ and $Z=10$. Here and below $\ell=1.8$, $a=0.35$, and $p=15$.}
\label{figure2}
\end{figure}

The linear Hamiltonian $H_0$ is $\ell-$periodic along the $y-$axis and $Z-$periodic along the $z-$axis. Defining $H=i\partial_t-H_0 $ we consider a solution of the linear problem $H\tpsi=0$ in the form of a Floquet state $\tpsi=e^{ibz}\phi$, where $\phi(\br,z)=\phi(\br,z+Z)$ and $b \in [-\omega/2,+\omega/2)$ with $\omega=2\pi/Z$, is a quasi-propagation-constant (tildes are used to emphasize that eigenstates are linear). Thus, $H\phi =-b \phi$ can be viewed as an eigenvalue problem, characterized by a Bloch spectrum $b=b_{\nu k} $, where  $k\in[-K/2,K/2)$ is the Bloch momentum along the $y$-axis in the reduced Brillouin zone (BZ), $K=2\pi/\ell$ is the width of the BZ, and $\nu$ enumerates allowed bands and defect states (if any) for a given $k$ {(corresponding Floquet state can also be written as $\tpsi_{\nu,k}$)}. The representative spectrum {(required for construction of the Floquet defect solitons and} calculated using propagation and projection method suggested in Ref.~\cite{edgesol03}) is illustrated in Fig.~\ref{figure2} for different numbers $M$ of the tilted layers.
%$M=1$~(a), $M=2$~(b) and $M=4$~(c). 
All these spectra feature an allowed band and clearly visible $M$ branches of the defect modes (localized in $x$ and extended in $y$ directions). {Such linear defect modes were recently observed in~\cite{defect01}}. Note that unlike edge states, usually confined to a boundary between two different media, a defect mode extends across the entire defect layer [see examples of all five linear defect modes supported by $M=5$ tilted layers in Fig.~\ref{figure2}(e)]. It is also noteworthy that Floquet dynamics is created only by mutual positioning of the layers of straight and tilted waveguides, rather than by helicity of each individual waveguide, considered in the most of previous studies.  The first $b^{\prime}=\partial b/\partial k$ and second $b^{\prime\prime}=\partial^2 b/\partial k^2$ derivatives of the quasi-propagation constant $b$ from Fig.~\ref{figure2} quantify averaged group velocity and dispersion of corresponding defect modes~\cite{edgesol06}.

The tilt angle of the defect layers is an important parameter defining the "frequency" of oscillations of the defect [illustrated in Fig.~\ref{figure1}(c)]. Formally, such angle can be made arbitrarily small, drawing the system to the adiabatic limit that corresponds to $\alpha\to 0$. Generally speaking, this limit is far from situation considered here, since in our case the period of oscillations of the defect is finite allowing one to use $z$-averaging~\cite{edgesol06,edgesol07} for derivation of the envelope equation describing Floquet defect solitons (see below, as well as {Appendix}). Meantime considered tilt angles are still sufficiently small, making it instructive instructive to consider the transition $\alpha\to 0$ from the point of view of the spectrum deformation resulting in a pure homogeneous array at $\alpha=0$ ($Z=\infty$). Transformation of the spectrum upon increase of $Z$ is shown in Fig.~\ref{figure2}(d) [cf. Fig.~\ref{figure2}(a)].

%%%%%%%%%%%%%%%%%%%%%%%%%%%%%%%%%%%%%%%%%%%%%%%%%%%%%%%%
%%%%%%%%%%%%%%%%%%%%%%%%%%%%%%%%%%%%%%%%%%%%%%%%%%%%%%%%

Suppose now that $\tpsi_{\nu k_0}(\br,z)$ is a linear defect mode, i.e., the respective field $\phi_{\nu k_0}$ solves  $H\phi_{\nu k_0}=-b_{\nu k_0}\phi_{\nu k_0}$ subject to the boundary conditions $\lim_{|x|\to\infty}\phi_{\nu k_0}(\br,z)=0$, and has the form of a Bloch wave along the $y$-direction: $\phi_{\nu k_0}=e^{ik_0x}u_{\nu k_0}$, where  $u_{\nu k_0}(\br,z)=u_{\nu k_0}(\br+\ell\hbj,z)=u_{\nu k_0}(\br,z+Z)$. The Bloch states can be chosen normalized as follows~\cite{edgesol06,edgesol07}:
$ \int_{-\infty}^\infty dx\int_{0}^\ell dy|u_{\nu  k_0} (\br,z)|^2 =1$ (note that this integral does not depend on $z$).
If nonlinearity is added, a branch of the nonlinear defect modes bifurcates from $\tpsi_{\nu k_0}(\br,z)$.  Respectively we look for a solution of (\ref{NLS_dimensionless}) using  the multiple-scale expansion, which for Floquet systems was elaborated in~\cite{edgesol06,edgesol07}. Here we omit the details and present only the final results. Since, the respective nonlinear modes are moving with velocity of the linear defect mode {(because they are constructed as envelopes for linear states from which they bifurcate)}, given in our case by  $-b_{\nu k}'$, the solution of Eq.~(\ref{NLS_dimensionless}) is searched in the form
%\begin{align}
%\label{psi_expan}
$
\psi= \mu e^{ib_{\nu k_0} z}[A(\eta, z)\phi_{\nu  k_0} (\br,z)   + \mu   \phi^{(1)}  
%+\mu^2  \phi^{(2)}
+\cdots],
$
%\end{align} 
where $\eta=y+b_{\nu k_0}'z$ is the "running" variable, and $A(\eta,z)$ is a slowly varying envelope of the defect state $\phi_{\nu k_0}$. The shape of the defect soliton is therefore described by $A(\eta,z)$ which solves the NLS equation:
 \begin{align}
     i\frac{\partial A}{\partial z}-\frac{b_{\nu k_0}''}{2}\frac{\partial^2 A}{\partial \eta^2}+\chi|A|^2A=0,
 \end{align} 
where $\chi=\frac{1}{Z}\int_0^Zdz\int_{-\infty}^\infty dx\int_{0}^\ell dy|\phi_{\nu  k_0} (\br,z)|^4 $ {(the derivation is outlined in Appendix)}.

%%%%%%%%%%%%%%%%%%%%%%%%%%%%%%%%%%%%%%%%%%%%%%%%%%%%%%%%
\begin{figure}[t]
\centering\includegraphics[width=1\linewidth]{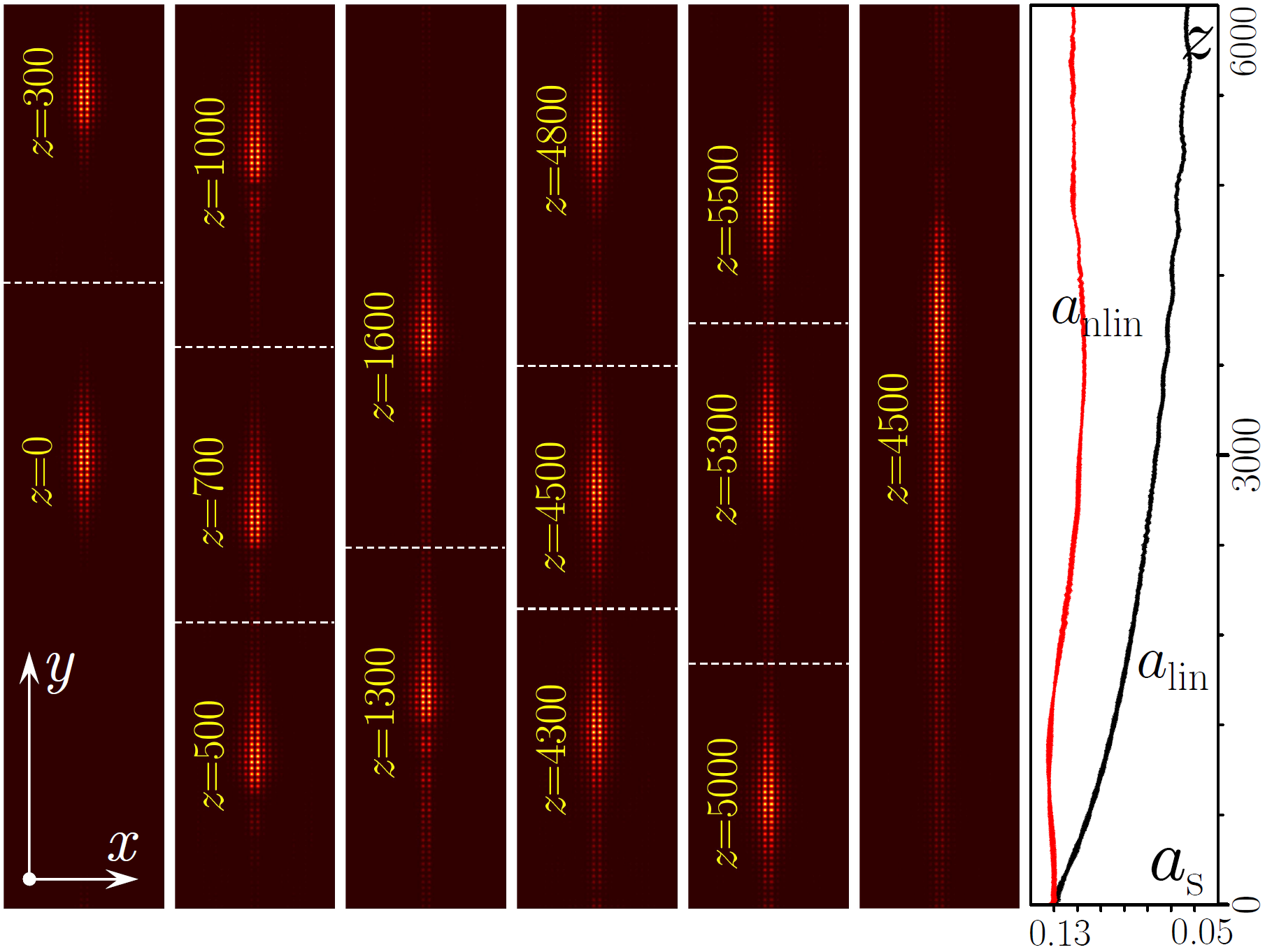}
\caption{Propagation of a defect soliton in structure with two tilted waveguide layers ($M=2$). Input envelope corresponds to $b_\textrm{nl}=0.003$, $k_0=0.3K$, $b_{2 k_0}^{\prime\prime}=-0.279$, $\chi=0.5116$. $|\psi|$ distributions at different distances are superimposed {(horizontal dashed lines are guides for an eye separating them)}, soliton moves in the positive $y$-direction. The right $|\psi|$ distribution shows output for linear medium at $z=4500$. The rightmost plot illustrates peak amplitude {$a_\textrm{s}$ versus distance $z$} for bright soliton launched into linear (black) and nonlinear (red) medium. {Here and below $Z=10$.}}
\label{figure3}
\end{figure}
%%%%%%%%%%%%%%%%%%%%%%%%%%%%%%%%%%%%%%%%%%%%%%%%%%%%%%%%

Since in our case $\chi>0$, the type of the defect soliton that can be excited in such system depends on the sign of $b_{\nu k_0}''$. To demonstrate the existence of bright solitons, we consider the photonic lattice sketched in the left panel of Fig.~\ref{figure1} with $M=2$ (in this case one can identify the defect levels in eigenvalue spectrum by $\nu=1$ and $\nu=2$). {Spectrum for this number of tilted layers has states with negative dispersion necessary for the formation of bright solitons - one of suitable states satisfying this requirement is indicated by the red dot in Fig. \ref{figure2}(b). Formation of solitons on defects with larger $M$ illustrating generality of the phenomenon is discussed below.} The input wavepacket is obtained by superimposing a bright envelope at $z=0$, i.e.,
$A(\eta,0)=(2b_\textrm{nl}/\chi)^{1/2} \mathrm{sech}[(-{2b_\textrm{nl}}/{b^{\prime\prime}_{\nu k_0} })^{1/2}\eta]$ corresponding to the nonlinearity-induced phase shift $b_\textrm{nl}$ of the propagation constant from quasi-propagation constant of the linear mode $b_{\nu k}$,
on the $\nu=2$ defect mode $\phi_{2 k_0}(\br,0)$ taken, for example, at $k_0=0.3K$, marked by the red dot in Fig.~\ref{figure2}(b), where $b_{1 k_0}=0.43\omega=0.27$ and $b_{2 k_0}^{\prime\prime}=-0.279$ (the dispersion is negative as required for the formation of bright solitons). Note that this mode can be viewed as an {\em embedded}~\cite{YangMalomed} defect soliton since its quasi-propagation-constant is within the allowed band of the Floquet spectrum. Fig.~\ref{figure3} illustrates propagation dynamics of the so-constructed defect bright soliton moving through the hundreds of $z$-periods. Distributions of $|\psi|$ at different propagation distances (indicated on the plots) are superimposed at each of the panels. One can see that the soliton propagates in the positive direction of the $y$-axis without noticeable changes in its shape, except for small oscillations behind it caused by higher-order dispersion, which is not taken into account in our theory. Such oscillations are visible only at the initial stage of the evolution of soliton and disappear at larger distances $z\geq4000$. Soliton has an amplitude that experiences small $Z$-periodic oscillations and on average remains almost constant in $z$.  

In the rightmost panel of Fig.~\ref{figure3} with $|\psi|$ distributions we show the result of propagation (output distribution) of the same input state, but in a linear medium. Unlike in the nonlinear case, after an initial transient interval the wavepacket exhibits strong asymmetric expansion and its peak amplitude substantially decreases. The comparison of peak amplitudes for linear $a_{\text{lin}}$ (black curve) and nonlinear $a_{\text{nlin}}$ (red curve) propagation regimes is illustrated in the rightmost panel of Fig.~\ref{figure3}.

%%%%%%%%%%%%%%%%%%%%%%%%%%%%%%%%%%%%%%%%%%%%%%%%%%%%%%%%
\begin{figure}[t]
\centering\includegraphics[width=0.94\linewidth]{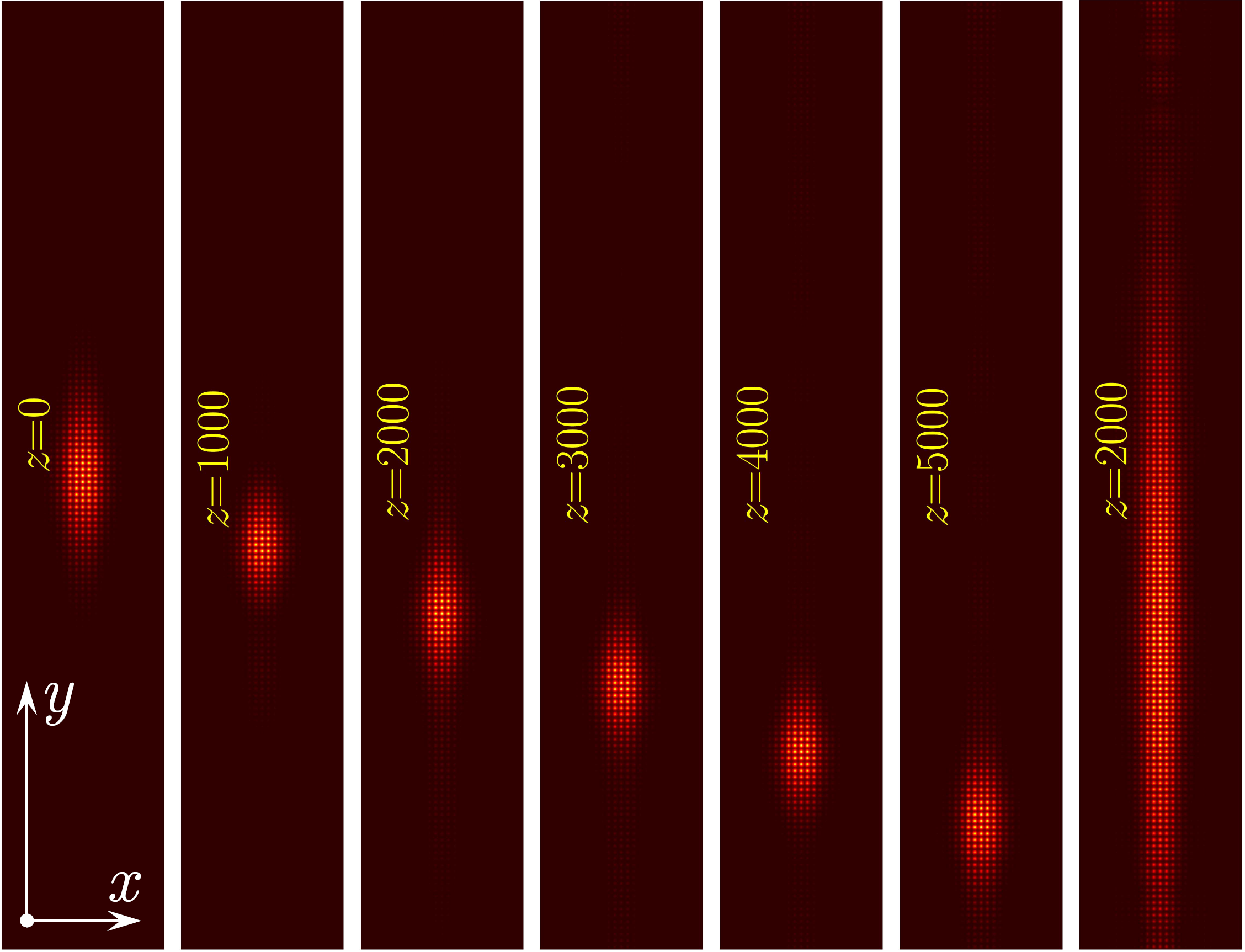}
\caption{Propagation a bright embedded defect soliton in structure with five {($M=5$)} tilted layers {(see Visualization 1)}. Input envelope corresponds to $b_\textrm{nl}=0.002$, $k_0=0.9K$, $b_{1 k_0}^{\prime\prime}=-0.3263$, $\chi=0.209$. Soliton moves in the negative $y$-direction. Right plot shows output for linear medium.}
\label{figure4}
\end{figure}
%%%%%%%%%%%%%%%%%%%%%%%%%%%%%%%%%%%%%%%%%%%%%%%%%%%%%%%%

Dynamics of bright defect soliton, bifurcating from the linear Floquet state for a defect with $M=5$ tilted layers at $k_0=0.9K$, shown in Fig.~\ref{figure2}(c) by the red dot with $b_{1 k_0}=0.028\omega=0.18$, is illustrated in Fig.~\ref{figure4}. Now the total quasi-propagation constant of the initial state is located in the gap of the linear spectrum. The defect soliton propagates over $500$ $z$-periods traversing about $55$ $y$-periods practically without changing its shape {and may even survive after interaction with missing channel, see Appendix}. An interesting feature of this evolution is that for these particular parameters the soliton moves in the negative direction of the $y$-axis (as predicted by the sign of $b_{1 k_0}^\prime$), although the waveguides of the defect layers are tilted in the positive $y-$direction. For comparison, the result of linear propagation for the same initial condition is shown in the rightmost panel of this figure.

%%%%%%%%%%%%%%%%%%%%%%%%%%%%%%%%%%%%%%%%%%%%%%%%%%%%%%%%
\begin{figure}[t]
\centering\includegraphics[width=0.94\linewidth]{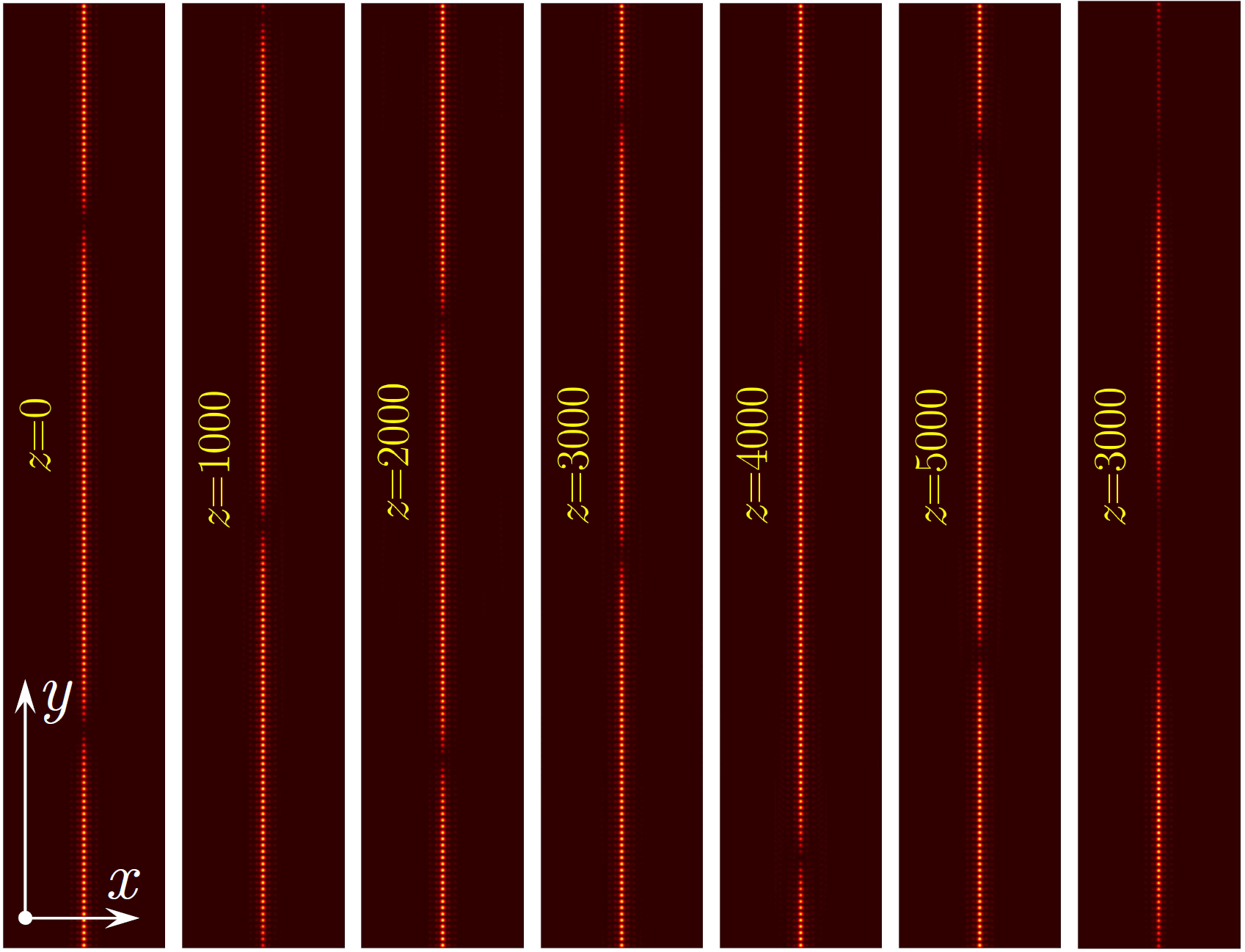}
\caption{Propagation of two spatially separated defect dark solitons in the array with one tilted layer {($M=1$)}. Input envelope corresponds to $k_0=0.6K$ [see Fig.~\ref{figure2}(a)], $b_\textrm{nl}=0.003$, $b_{1 k_0}^{\prime\prime}=0.5158$, $\chi=1.1369$. Soliton moves in the positive $y$-direction. Right plot shows output for the linear medium.}
\label{figure5}
\end{figure}
%%%%%%%%%%%%%%%%%%%%%%%%%%%%%%%%%%%%%%%%%%%%%%%%%%%%%%%%

Since the Floquet-Bloch spectrum of the linear defect modes has intervals with positive dispersion, the described arrays can also support Floquet defect dark solitons with the respective envelope $A(\eta,0)=(b_\textrm{nl}/\chi)^{1/2} \tanh[({b_\textrm{nl}}/{b^{\prime\prime}_{\nu k_0} })^{1/2}\eta]$. An example of propagation of a dark soliton for $M = 1$ is shown in the Fig.~\ref{figure5}. To ensure field periodicity on a large, but finite, $y$-window, a pair of well-separated initially located in the points $\eta=\pm 40\ell$ identical dark solitons with opposite phases is simultaneously nested on the Bloch wave. Such solitons bifurcate from the branch at $k_0=0.6K$ with positive dispersion $b_{1 k_0}^{\prime\prime}=0.5158$ and $b_{1 k_0}=0.07\omega=0.04$ shown in Fig.~\ref{figure2}(a) by the red dot. No signs of the background instability are visible even at distances $z \sim 5000$ and there is almost no radiation into the bulk of the lattice. In contrast, the same state strongly disperses in the linear medium, as visible in the corresponding output in the rightmost panel of Fig.~\ref{figure5}.

Summarizing, we predicted that $z$-periodic defects created by several tilted layers of waveguides embedded in otherwise uniform square waveguide arrays can support robust Floquet defect solitons of both bright and dark types. {These hybrid strongly asymmetric states inherit localization in one spatial direction from linear defect modes and are localized due to self-action in other direction. For this reason they do not feature power threshold for their excitation and can be encountered in various Floquet systems with different symmetries.} 

\medskip

The authors acknowledge funding from the RSF (grant  21-12-00096), and Portuguese Foundation for Science and Technology (FCT) under Contract no. UIDB/00618/2020.

\medskip

\noindent\textbf{Disclosures.} The authors declare no conflicts of interest.

% Bibliography
%\bibliography{sample}

\begin{thebibliography}{99}

\bibitem{dinlattices01}
I. L. Garanovich, S. Longhi, A. A. Sukhorukov, and Y. S. Kivshar,
"Light propagation and localization in modulated photonic lattices and waveguides,"
Phys. Rep. \textbf{518}, 1 (2012).

\bibitem{topophot02}
T. Ozawa, H. M. Price, A. Amo, N. Goldman, M. Hafezi, L. Lu, M. C. Rechtsman, D. Schuster, J. Simon, O. Zilberberg and I. Carusotto,
"Topological photonics,''
Rev. Mod. Phys. \textbf{91}, 015006 (2019).

\bibitem{floquet04}
N. Goldman and J. Dalibard,
"Periodically driven quantum systems: effective Hamiltonians and engineered gauge fields,"
Phys. Rev. X \textbf{4}, 031027 (2014).
 
\bibitem{floquet02}
M. S. Rudner, N. H. Lindner,
"Band structure engineering and nonequilibrium dynamics in Floquet topological insulators,”
Nat. Rev. Phys. \textbf{2}, 229 (2020).

%\bibitem{floquet03}
%N. Goldman, G. Juzeliūnas, P. Öhberg, and I. B. Spielman,
%"Light-induced gauge fields for ultracold atoms,"
%Rep. Prog. Phys. \textbf{77}, 126401 (2014).

\bibitem{topophot01}
L. Lu, J. D. Joannopoulos and M. Solja{\v{c}}i\'c,
"Topological photonics,''
Nat. Photon. \textbf{8}, 821 (2014).

 
\bibitem{floquet05}
N. H. Lindner, G. Refael, and V. Galitski,
"Floquet topological insulator in semiconductor quantum wells,"
Nat. Phys. \textbf{7}, 490 (2011).

\bibitem{floquet06}
M. S. Rudner, N. H. Lindner, E. Berg, and M. Levin,
"Anomalous edge states and the bulk-edge correspondence for periodically driven two-dimensional systems,"
Phys. Rev. X \textbf{3}, 031005 (2013).

\bibitem{floquet07}
M. C. Rechtsman, J. M. Zeuner, Y. Plotnik, Y. Lumer, D. Podolsky, F. Dreisow, S. Nolte, M. Segev, and A. Szameit,
"Photonic Floquet topological insulators,''
Nature \textbf{496}, 196 (2013).

\bibitem{floquet08}
L. J. Maczewsky, J. M. Zeuner, S. Nolte, and A. Szameit,
"Observation of photonic anomalous Floquet topological insulators,"
Nat. Commun. \textbf{8}, 13756 (2017).

\bibitem{floquet09}
S. Mukherjee, A. Spracklen, M. Valiente, E. Andersson, P. Öhberg, N. Goldman, R. R. Thomson,
"Experimental observation of anomalous topological edge modes in a slowly driven photonic lattice,"
Nat. Commun. \textbf{8}, 13918 (2017).

\bibitem{edgesol01}
Y. Lumer, Y. Plotnik, M. C. Rechtsman, and M. Segev,
"Self-localized states in photonic topological insulators,''
Phys. Rev. Lett. \textbf{111}, 243905 (2013).

\bibitem{edgesol02}
M. J. Ablowitz, C. W. Curtis, and Y.-P. Ma,
"Linear and nonlinear traveling edge waves in optical honeycomb lattices,''
Phys. Rev. A \textbf{90}, 023813 (2014).

\bibitem{edgesol03}
D. Leykam and Y.-D. Chong,
"Edge solitons in nonlinear-photonic topological insulators,''
Phys. Rev. Lett. \textbf{117}, 143901 (2016).

\bibitem{edgesol04}
M. J. Ablowitz and J. T. Cole,
"Tight-binding methods for general longitudinally driven photonic lattices: Edge states and solitons,''
Phys. Rev. A \textbf{96}, 043868 (2017).

\bibitem{edgesol05}
S. K. Ivanov, Y. V. Kartashov, A. Szameit, L. J. Maczewsky, and V. V. Konotop,
"Edge solitons in Lieb topological Floquet insulators,''
Opt. Lett. \textbf{45}, 1459 (2020).

\bibitem{edgesol06}
S. K. Ivanov, Y. V. Kartashov, A. Szameit, L. Torner, V. V. Konotop,
"Vector topological edge solitons in Floquet insulators,''
ACS Photonics \textbf{7}, 735 (2020).

\bibitem{edgesol07}
S. K. Ivanov, Y. V. Kartashov, M. Heinrich, A. Szameit, L. Torner, and V. V. Konotop,
"Topological dipole Floquet solitons,''
Phys. Rev. A \textbf{103}, 053507 (2021).

\bibitem{edgesol08}
S. K. Ivanov, Y. V. Kartashov, L. J. Maczewsky, A. Szameit, V. V. Konotop,
"Bragg solitons in topological Floquet insulators,''
Opt. Lett. \textbf{45}, 2271 (2020).

\bibitem{edgesol09}
D. A. Smirnova, L. A. Smirnov, D. Leykam, Y. S. Kivshar,
"Topological edge states and gap solitons in the non-linear Dirac model,"
Laser Photon. Rev. \textbf{13}, 1900223 (2019). 

\bibitem{edgesol10}
Z. Shi, M. Zuo, H. Li, D. Preece, Y. Zhang, and Z. Chen,
"Topological edge states and solitons on a dynamically tunable domain wall of two opposing helical waveguide arrays,"
ACS Photonics \textbf{8}, 1077 (2021).

\bibitem{solobsv01}
S. Mukherjee, M. C. Rechtsman,
"Observation of Floquet solitons in a topological bandgap,''
Science \textbf{368}, 856 (2020).

\bibitem{solobsv02}
S. Mukherjee, M. C. Rechtsman,
"Observation of unidirectional soliton-like edge states in nonlinear Floquet topological insulators,”
arXiv:2010.11359 (2020).

\bibitem{solobsv03}
L. J. Maczewsky, M. Heinrich, M. Kremer, S. K. Ivanov, M. Ehrhardt, F. Martinez, Y. V. Kartashov, V. V. Konotop, L. Torner, D. Bauer, and A. Szameit,
"Nonlinearity-induced photonic topological insulator,"
Science \textbf{370}, 701 (2020).

\bibitem{nonlrev01}
D. Smirnova, D. Leykam, Y. D. Chong, and Y. Kivshar,
"Nonlinear topological photonics,''
Appl. Phys. Rev. \textbf{7}, 021306 (2020).

\bibitem{curved01}
I. L. Garanovich, A. A. Sukhorukov, and Y. S. Kivshar,
"Defect-free surface states in modulated photonic lattices,"
Phys. Rev. Lett. \textbf{100}, 203904 (2008).

\bibitem{curved02}
A. Szameit, I. L. Garanovich, M. Heinrich, A. A. Sukhorukov, F. Dreisow, T. Pertsch, S. Nolte, A. Tunnermann, and Y. S. Kivshar,
"Observation of defect-free surface modes in optical waveguide arrays,"
Phys. Rev. Lett. \textbf{101}, 203902 (2008).

\bibitem{curved03}
S. Longhi and G. Della Valle,
"Floquet bound states in the continuum,"
Sci. Rep. \textbf{3}, 2219 (2013).

\bibitem{drivenaah01}
Y. G. Ke, X. Z. Qin, F. Mei, H. H. Zhong, Yu. S. Kivshar, and C. Lee,
"Topological phase transitions and Thouless pumping of light in photonic waveguide arrays,"
Las. Photon. Rev. \textbf{10}, 995 (2016).

\bibitem{drivenssh01}
J. K. Asb\'oth, B. Tarasinski, P. Delplace,
"Chiral symmetry and bulk-boundary correspondence in periodically driven one-dimensional systems,"
Phys. Rev. B \textbf{90}, 125143 (2014).

\bibitem{drivenssh02}
V. Dal Lago, M. Atala, L. E. F. Foa Torres,
"Floquet topological transitions in a driven one-dimensional topological insulator,"
Phys. Rev. A \textbf{92}, 023624 (2015).

\bibitem{drivenssh03}
Y. Zhang, Y. V. Kartashov, F. Li, Z. Zhang, Y. Zhang, M. R. Beli\'c, and M. Xiao,
"Edge states in dynamical superlattices,"
ACS Photonics \textbf{4}, 2250 (2017).

\bibitem{drivenssh04}
B. Zhu, H. Zhong, Y. Ke, X. Qin, A. A. Sukhorukov, Y. S. Kivshar, and C. Lee,
"Topological Floquet edge states in periodically curved waveguides,"
Phys. Rev. A \textbf{98}, 013855 (2018).

\bibitem{drivenssh05}
Q. Cheng, Y. Pan, H. Wang, C. Zhang, D. Yu, A. Gover, H. Zhang, T. Li, L. Zhou, S. Zhu,
"Observation of anomalous $\pi$ modes in photonic Floquet engineering,"
Phys. Rev. Lett. \textbf{122}, 173901 (2019).

\bibitem{defect01}
Y. Lumer, M. A. Bandres, M. Heinrich, L. J. Maczewsky, H. Herzig-Sheinfux, A. Szameit, and M. Segev,
"Light guiding by artificial gauge fields,''
Nat. Photon. \textbf{13}, 339 (2019).

\bibitem{defect02}
W. Song, Y. Chen, H. Li, S. Gao, S. Wu, C. Chen, S. Zhu, and T. Li,
"Gauge-induced Floquet topological states in photonic waveguides,"
Las. Photon. Rev. \textbf{15}, 2000584 (2021).

\bibitem{YangMalomed} J. Yang, B. A. Malomed, and D. J. Kaup, 
"Embedded solitons in second-harmonic-generating systems,"
Phys. Rev. Lett. \textbf{83}, 1958 (1999).



\end{thebibliography}
% Full bibliography added automatically for Optics Letters submissions; the following line will simply be ignored if submitting to other journals.
% Note that this extra page will not count against page length
%\bibliographyfullrefs{sample}

\appendix

\section{Derivation of the nonlinear Schr\"odinger equation (2).}

Here we outline the derivation of the NLS equation (2) from the main text. Some details of the derivation procedure similar to those described in details in earlier publications~\cite{edgesol06,edgesol07}, are not reproduced here. 

We start with the main nonlinear parabolic equation~(1)
\begin{align}  
\label{main}
i\frac{\partial \psi}{\partial z}=-\frac{1}{2}\nabla^2 \psi-\mathcal{R}(\br,z)\psi-|\psi|^2\psi 
\end{align}
We use the same notations as in the main text, i.e.,   $\nabla=(\partial_x,\partial_y)$, ${\bm r}=(x,y)$,
\begin{align}
\mathcal{R}(\br)=R_0(\br)+R_d(\br,z)
\end{align}
with
\begin{align}
R_0(\br)=R_0(\br+\ell\hbi)=R_0(\br+\ell\hbj)
\end{align}
term that describes perfect periodic 2D square lattice and
\begin{align}
R_d(\br,z)=\begin{cases}
R_0(\br-\alpha z\hbj)-R_0(\br), & |x|<L/2
\\
0, & |x|>L/2
\end{cases}
\end{align}
term that describes the \textit{defect} (tilted layer).
Thus:
\begin{eqnarray}
\mathcal{R}(\br,z)=\mathcal{R}(\br+\ell\hbj,z)= \mathcal{R}(\br,z+Z)
\end{eqnarray}

Consider the linear problem
\begin{align}  
\label{main_lin}
i\frac{\partial \tpsi}{\partial z}=-\frac{1}{2}\nabla^2 \tpsi-\mathcal{R}(\br,z)\tpsi 
\end{align}
and the ansatz
\begin{align}
\tpsi(\br,z)=e^{ibz}\phi(\br,z), \qquad  b\in\left[-\frac{\pi}{Z},\frac{\pi}{Z} \right)
\end{align}
where the function $\phi(\br,z)$ is periodic along the $z-$axis 
\begin{align}
\phi(\br,z)=\phi(\br,z+Z)
\end{align}
and solves the equation
\begin{align}  
\label{main_phi}
i\frac{\partial \phi}{\partial z}=-\frac{1}{2}\nabla^2 \phi-\mathcal{R}(\br,z)\phi+b\phi 
\end{align}
Next we use the fact that due to periodicity of the array along the $y-$axis, the function $\phi(x,y,z)$ can be searched in the form of a Bloch wave:
\begin{align}
\label{Bloch_funct}
\phi(\br,z)=e^{iky}u_{\nu k}(\br,z),   
\end{align}
where
\begin{align}
u_{\nu k}(\br,z)=u_{\nu k}(\br+\ell\hbj,z)=u_{\nu k}(\br,z+Z), 
\end{align}
with Bloch momentum $k$ in the first Brillouin zone
\begin{align}
k	\in\left[-\frac{\pi}{\ell},\frac{\pi}{\ell} \right)
\end{align}
and with index $\nu$ enumerating modes at a given value of $k$ (in the absence of the level crossing). 

Finally, we turn to the $x-$dependence of the solution. For a mode to be localized along $x$ axis in the defect (tilted) layer, we have to impose zero boundary conditions
\begin{align}
\lim_{x\to\pm \infty} u_{\nu k}(\br,z) =0 
\end{align}
Since at $|x|>L/2$ the lattice is unperturbed, in these domains one can look for a solution in the form 
\begin{align}
u_{\nu k}(\br,z)= e^{-|\lambda| x}w_{\nu k}(\br,z) 
\end{align} 
where
\begin{equation}
\begin{split}
w_{\nu k}(\br,z)=&w_{\nu k}(\br+\ell\hbi,z)= \\
\qquad &w_{\nu k}(\br+\ell\hbj,z)=w_{\nu k}(\br,z+Z),
\end{split}
\end{equation}
$\lambda$ is a constant and we used the fact that the system is symmetric with respect to the inversion $x\to-x$ considering the decay at $x\to-\infty$ and at $x\to \infty$ characterized by the same exponent.

We are interested in the simplest soliton solutions, which are constructed (bifurcate from) on linear defect modes localized in the $x$-direction. Taking into account that in realistic applications one deals with finite structures we solve the eigenvalue problem (\ref{main_phi}) on sufficiently large, but finite $x$-window of width $X$, using plane-wave expansion method. Even though this method assumes periodic boundary conditions:
\begin{align}
u_{\nu k}(x-X/2,y,z)=u_{\nu k}(x+X/2,y,z)
\end{align}
they are nevertheless fully consistent with exponential localization of the defect modes, since the latter have amplitude of the order of $10^{-9}$ at the borders of the integration domain. For the parameters used in the present work and for selected $k$ values we have found the number of localized modes equal to the number of the tilted layers in the defect. These defect modes coexist with bulk modes that extend far beyond the defect, into the bulk of stationary array. The solution of the linear eigenvalue problem (\ref{main_phi}) eventually yields the Floquet exponent $b$ as a function of $\nu$ and $k_0$ (see (\ref{Bloch_funct})), that is depicted in the panels (a)-(d) of Fig.~2 of the main text.

A quasi-one-dimensional envelope soliton can be constructed by imposing broad envelope on a mode localized in the defect region and having a Bloch wave number $k_0$. The procedure of the multiple-scale expansion combined with the averaging over $Z$-period is described in all details in Refs.~\cite{edgesol06,edgesol07}. Therefore, here we outline only the main steps. 

Supposing that the soliton is constructed on the mode with index $\nu$, we perform the multiscale expansion by introducing  a formal small parameter $0<\mu\ll 1$, two sets of formally independent scaled variables $(y_0,y_1,y_2,...):=(y, \mu y, \mu^2 y,...)$ and $(z_0,z_1,z_2...):=(z, \mu z, \mu^2 z,...)$, and look for a solution of Eq.~(\ref{main}) in the form of the expansion 
\begin{equation}
\begin{split}
\label{ansatz}
&\psi= e^{ib_{\nu k_0} z_0}\phi,\\ &\phi=\mu  A(\eta, z_2)\phi_{\nu  k_0} (\br,z_0)   + \mu^2   \phi^{(1)}  + \mu^3   \phi^{(2)} +\cdots
\end{split}  
\end{equation}
where $\eta=y_1+b_{\nu k_0}'z_1$: $A(\eta, z)=A(y_1+b_{\nu k_0}' z_1, z_2)$ [the convention that in the arguments of the amplitude only the slowest variables are indicated is used]. The Bloch state $\phi_{\nu k_0}$ depends only on the "fast" variables $(x,y_0,z_0)$; the variable $x$ is not scaled. All slow variables are considered as independent.

Defining the linear operator
\begin{equation}
\begin{split}
\label{H0}
&H_0:= i\frac{\partial}{\partial z_0}+\frac{1}{2}\left(\frac{\partial^2 }{\partial x^2}+\frac{\partial^2 }{\partial y_0^2}\right) + \mathcal{R}(\br_0,z)+b_{\nu k_0}, \\ &\br_0=(x,y_0)
\end{split}
\end{equation}
and using that in the slow variables
\begin{align}
\frac{\partial}{\partial y}=\frac{\partial}{\partial y_0}+\mu\frac{\partial}{\partial y_1}+\mu^2\frac{\partial}{\partial y_2}+\cdots
\\
\frac{\partial}{\partial z}=\frac{\partial}{\partial z_0}+\mu\frac{\partial}{\partial z_1}+\mu^2\frac{\partial}{\partial z_2}+\cdots
\end{align}
we substitute (\ref{ansatz}) into (\ref{main}) and collect the terms with different orders of $\mu$. This gives (only the three leading orders are shown):
\begin{equation}
\begin{split}
\label{line1}
0&= A H_0\phi_{\nu  k_0}  
\\
&+\mu \left(H_0 \phi^{(1)}+i\frac{\partial A}{\partial z_1}
\phi_{\nu  k_0}+\frac{\partial \phi_{\nu  k_0}}{\partial y_0}\frac{\partial A}{\partial y_1}\right)
\\
&+\mu^2 \left(H_0 \phi^{(2)}+i\frac{\partial A}{\partial z_2}
\phi_{\nu  k_0}+\frac{\partial \phi_{\nu  k_0}}{\partial y_0}\frac{\partial A}{\partial y_2}+\frac{\partial^2 \phi^{(1)}}{\partial y_0\partial y_1}+ \right.
\\
&\qquad\qquad \left.\frac{1}{2}\frac{\partial^2 A}{\partial y_1^2}\phi_{\nu  k_0}+|A|^2A|\phi_{\nu  k_0}|^2\phi_{\nu  k_0}\right)
\end{split}
\end{equation}
The expressions in each of the lines in this equation must be zero independently. The ansatz (\ref{ansatz}) ensures that the leading order in the first line of (\ref{line1}) is zero identically. In order to resolve the orders of $\mu$ in the second line of (\ref{line1}) and $\mu^2$ in the third line of (\ref{line1}) we expand $\phi^{(1,2)}$ over the complete set of eigenfunctions ($j=1,2$)
\begin{align}
\label{expan_first}
\phi^{(j)} =\sum_{\nu'} C_{\nu' k_0}^{(j)}(y_1,z_0)\phi_{\nu'k_0}(y_0,z_0)
\end{align}
where, as explained in~\cite{edgesol06,edgesol07}, $C_{\nu' k_0}^{(j)}(y_1,z_0)=C_{\nu'k_0}^{(j)}(y_1,z_0+Z)$ and it is enough to account only the eignefunctions of the different bands with the same $k_0$. 

To proceed we define time averaging  over the period $Z$: 
\begin{align}
\langle f\rangle_T=\frac{1}{Z}\int_{0}^{Z}f(\br,z)dz    
\end{align}
and the inner product
\begin{align}
\label{inner}
(f,g)=\int_{-\infty}^{\infty} dx\int_0^\ell dy f^*(\br,z) g(\br,z). 
\end{align}

Equating the second line of (\ref{line1}) to zero we obtain~\cite{edgesol06}
\begin{equation}
\begin{split}
\label{mu2order_1}
i\frac{\partial A}{\partial z_1}\phi_{\nu k_0}+&\frac{\partial A}{\partial y_1}\frac{\partial \phi_{\nu k_0}}{\partial y_0} 
%\nonumber \\
=\\
&\sum_{\nu'}\left[\frac 1i \frac{  \partial C_{\nu'k_0}^{(1)} }{\partial z_0}  + (b_{\nu k_0}-b_{\nu' k_0})C_{\nu' k_0}^{(1)}\right]\phi_{\nu'k_0}.
\end{split}
\end{equation}
Taking into account that $A$ depends on $y_1$ and $z_1$ through $\eta$, where the relation 
\begin{align}
\label{omega1}
b_{\nu k_0}^\prime = \left\langle \left(\phi_{\nu k_0},i\frac{\partial \phi_{\nu k_0}}{\partial y_0} \right)\right\rangle_Z  
\end{align}
is obtained by the generalized $k\cdot p$ method described in~\cite{edgesol06,edgesol07}, one can ensure that application of $\langle (\phi_{\nu k_0},\cdot)\rangle_Z$  to the secon line of (\ref{line1}) yields zero. On the other hand, by applying to the same line $\langle (\phi_{\nu' k_0},\cdot)\rangle_Z$ with $\nu'\neq\nu $, one can obtain that $ C_{\nu'k_0}^{(1)} \propto  \partial A/\partial y_0$ (the proportionality coefficients can be found in ~\cite{edgesol06}.

The generalized $k\cdot p$ method also allows to express $b_{\nu k_0}^{\prime\prime}$ through the Floquet-Bloch eigenstates~\cite{edgesol06}:
\begin{equation}
\begin{split}
\label{dispersion}
\frac{1}{2}\frac{\partial^2 A}{\partial y_1^2}-&i\left\langle \sum_{\nu'\neq \nu} \left(\phi_{\nu k_0},i\frac{\partial \phi_{\nu' k_0}}{\partial y_0} \right) 
%(\phi_{\nu k_0},i\partial_{y_0}\phi_{\nu' k_0})
\frac{\partial C_{\nu'k_0}^{(1)} }{\partial y_1} \right\rangle_Z
=\\
&-\frac{b_{\nu k_0}^{\prime\prime}}{2}\frac{\partial^2 A}{\partial \eta^2}
\end{split}
\end{equation}
Now, equating the third line of (\ref{line1}) to zero, applying to it $\langle (\phi_{\nu k_0},\cdot)\rangle_Z$, using the relations (\ref{omega1}) and (\ref{dispersion}), employing the Fredholm alternative (i.e., eliminating secular terms), and assuming that the envelope $A$ is independent of the slow variable $y_2$, we arrive at the effective NLS equation (2) from the main text, where $\mu$ was set to one~\cite{edgesol06,edgesol07}.

\section{Interaction with a defect}

%%%%%%%%%%%%%%%%%%%%%%%%%%%%%%%%%%%%%%%%%%%%%%%%%%%%%%%%
\begin{figure*}[t]
	\centering\includegraphics[width=0.9\linewidth]{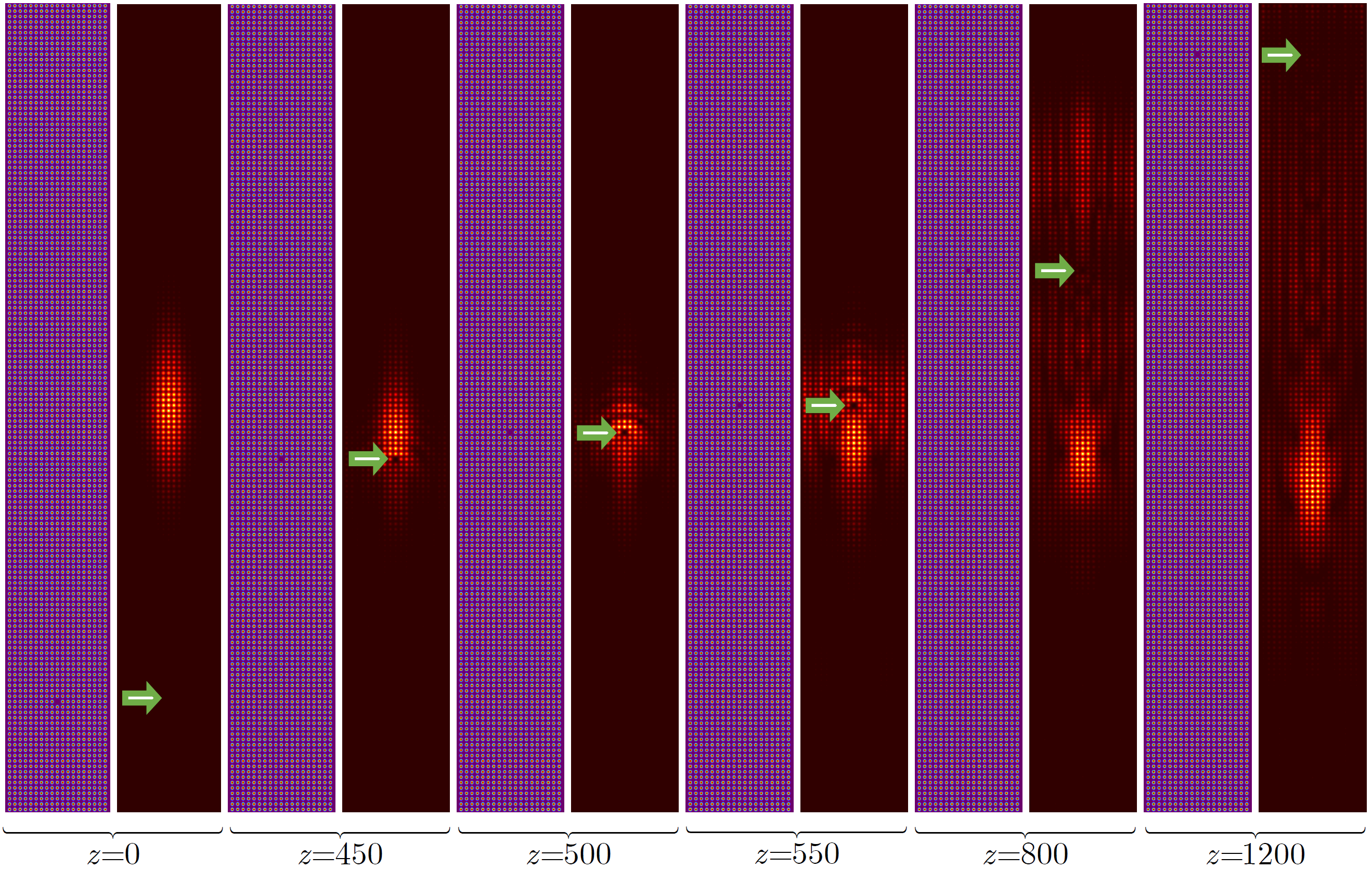}
	\caption{Interaction of soliton from Fig.~4 of the main text with the defect in the form of missing waveguide (see Visualization 2). Profiles of the array and field modulus distributions are shown at the same distances $z$. Green arrows indicate instantaneous defect position. }
	\label{figureSup}
\end{figure*}
%%%%%%%%%%%%%%%%%%%%%%%%%%%%%%%%%%%%%%%%%%%%%%%%%%%%%%%%

To study the impact of defects in the underlying structure on soliton propagation we consider the interaction of a bright soliton with a defect in the form of missing waveguide in a tilted layer. As an input, we use the same soliton as in Fig. 4 from the main text of the Letter with $M=5$. Input soliton is located at the point $y = 0$, while the missing waveguide is initially located far from soliton, at the point $y = -99$ (see left panels in Fig.~\ref{figureSup}). The figure shows field modulus distributions at different distances, when soliton only slightly touches the defect (second set of panels), when it fully overlaps with it (third set of panels), and when defect is passed (fourth to sixths sets of panels). One can see that even though the interaction with the defect causes notable backscattering and radiation into the bulk of the structure, the soliton survives after collision with defect and only its peak amplitude slightly decreases.
%Similar picture was encountered for dark solitons that also can bypass the defect.

\end{document}